\documentclass[aps,prx,print,groupedaddress]{revtex4-2}
\usepackage[utf8]{inputenc}
\usepackage{amsmath}
\usepackage{amssymb}
\usepackage{cancel}
\usepackage{bbold}
\usepackage{booktabs}
\usepackage{tabstackengine}
\usepackage{graphicx}
\usepackage{bm}
\usepackage{xcolor}
\usepackage{color,soul}

\usepackage{hyperref}
\hypersetup{%
	pdftitle={Application of deep neural networks for computing the renormalization group flow of the two-dimensional phi-4 field theory},%
	pdfauthor={Yueqi Zhao et al.},%
	pdfpagemode={UseNone},%
	pdfstartview={FitH},%
	breaklinks=true,%
	citecolor=blue,%
	colorlinks=true,%
	linkcolor=blue,%
	urlcolor=blue
}

\stackMath

\DeclareMathOperator{\Tr}{Tr}
\DeclareMathOperator{\arccosh}{arccosh}
\newcommand{\vect}[1]{{\bm{#1}}}

\begin{document}

\title{Application of deep neural networks for computing the renormalization group flow of the two-dimensional \texorpdfstring{$\boldsymbol{\phi}^4$}{phi-4} field theory}

\author{Yueqi Zhao}
\affiliation{Department of Physics, University of California San Diego, La Jolla, CA 92093, USA}

\author{Michael M. Fogler}
\affiliation{Department of Physics, University of California San Diego, La Jolla, CA 92093, USA}

\author{Yi-Zhuang You}
\affiliation{Department of Physics, University of California San Diego, La Jolla, CA 92093, USA}

\begin{abstract}
    We introduce RGFlow, a deep neural network–based real-space renormalization group (RG) framework tailored for continuum scalar field theories. Leveraging generative capabilities of flow-based neural networks, RGFlow autonomously learns real-space RG transformations from data without prior knowledge of the underlying model. In contrast to conventional approaches, RGFlow is bijective (information-preserving) and is optimized based on the principle of minimal mutual information. We demonstrate the method on two examples. The first one is a one-dimensional Gaussian model, where RGFlow is shown to learn the classical decimation rule. The second is the two-dimensional $\phi^4$ theory, where the network successfully identifies a Wilson–Fisher-like critical point and provides an estimate of the correlation-length critical exponent.
\end{abstract}

\maketitle

\section{Introduction}
\label{sec:introduction}

\begin{figure}
    \centering
    \includegraphics[width=3.5in]{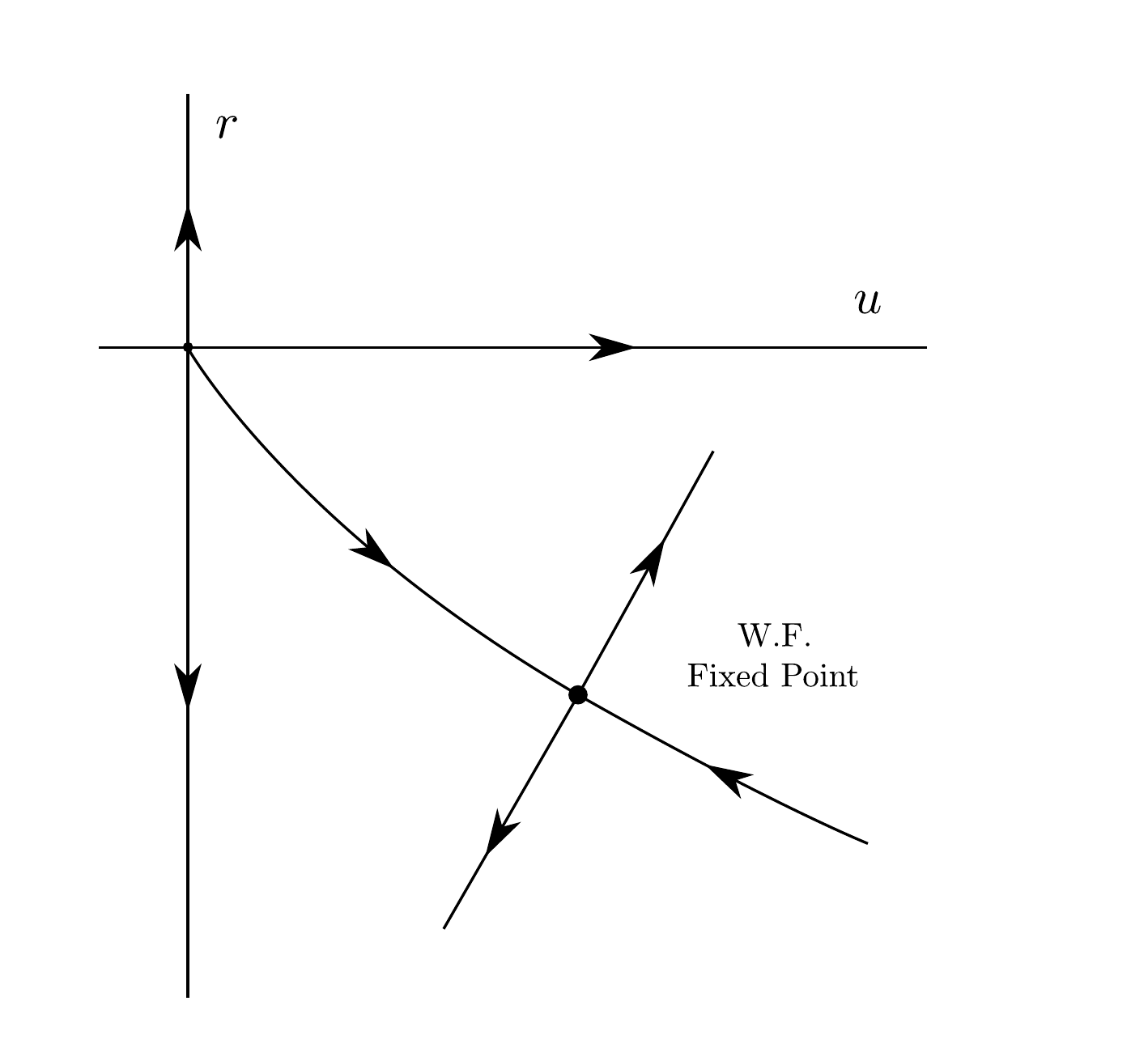}
    \caption{A cartoon drawing of the Wilson-Fisher fixed point of the $\phi^4$ model in $d = 4 - \epsilon$ dimension. The arrows indicate the directions of the RG flow.
    For the definition of the model and its parameters $r$, $u$, see Eq.~\eqref{eqn:8}.
    }
    \label{fig:W_F_fixed_point}
\end{figure}

The renormalization group (RG) framework is one of the cornerstones of modern theoretical physics, a powerful tool for analyzing systems with interacting degrees of freedom. It is typically formulated for effective actions of the form $\mathcal{S}(\textbf{K})$, where $\textbf{K}$ denotes a set of coupling constants and the theory includes an ultraviolet (UV) cutoff.
In this work, we consider real-space RG transformations, in which the cutoff corresponds to the minimal spatial scale---such as the nearest-neighbor distance $a$ on a lattice. The RG formalism describes how the parameters $\textbf{K}$ evolve as this scale is varied. Over the decades, a range of techniques have been devised to implement such transformations. Among them, the block decimation method for spin systems is widely used: the system is divided into local clusters, and rules like majority voting or simple averaging are applied within each block to define new coarse-grained variables. This leads to an RG map $\textbf{K}' = \mathcal{R}(\textbf{K})$ and an effective rescaling of the lattice constant to $a' = b a$, with $b > 1$. Of particular interest are the fixed points of this map, which correspond to phases if these fixed points are stable or phase transitions if they are unstable.
For example, the celebrated Wilson-Fisher fixed point (Fig.~\ref{fig:W_F_fixed_point}) describes the Ising-like ferromagnetic transition of the $\phi^4$ field theory.

However, traditional RG schemes have some limitations. Their effectiveness often depends on the choice of block structure and averaging rules. While increasing cluster size can improve accuracy, it also makes the RG transformation more complex and analytically intractable. Moreover, designing an RG flow that captures the key physics requires significant human intuition --- one must often guess the relevant order parameters in advance and construct rules tailored to them. For example, the na\"ive block-spin majority-vote RG scheme does not directly apply to antiferromagnetic systems, as the coarse-graining rule must be adapted to account for the underlying antiferromagnetic order in order to preserve the relevant ordering at larger scales. Therefore, it is desirable to develop automated, data-driven RG approaches that do not rely on human intuition to construct coarse-graining rules, but can instead learn them directly from microscopic configurations sampled from models at different length scales.

Recently, deep neural networks (DNNs) have emerged as powerful tools for automating complex tasks across a wide array of scientific disciplines. Applications range from data interpretation and feature extraction in microscopy~\cite{ziatdinov_deep_2017,gordon_scanning_2019,borodinov_deep_2019,lee_deep_2020,chen_hybrid_2021,zhao_deep-learning-aided_2023-1} to the design and optimization of photonic or electronic devices~\cite{hammond_designing_2019,gordon_automated_2020,chen_physics-informed_2020,dong_cktgnn_2024}. In statistical physics, DNNs have also been applied to design effective RG transformations at the configuration level~\cite{koch-janusz_mutual_2018,li_neural_2018,lenggenhager_optimal_2020,hu_machine_2020-1,chung_neural_2021,ron_monte_2021,giataganas_neural_2022,bachtis_inverse_2022,sheshmani_categorical_2023}. Yet for the exception of the systems with descrete variables (spins)~\cite{di_sante_deep_2022,ueda_finite-size_2023,hou_machine_2023}, relatively few studies have tackled the direct computation of the RG flow. 

In this work, we present RGFlow, a DNN-based real-space RG framework tailored to continuum scalar field theories. Our approach learns transformation rules from data and generates RG flow diagrams that encapsulate the model's fixed points and critical exponents.
The structure of this paper is as follows. Section~\ref{sec:architecture} introduces the architecture of RGFlow and details the network design. In Section \ref{sec:analytical}, we consider the RGFlow optimization problem for a one-dimensional Gaussian model, which is analytically solvable. Section~\ref{sec:results} demonstrates the application of RGFlow to the two-dimensional $\phi^4$ theory, revealing a Wilson–Fisher-like critical point (Fig. \ref{fig:W_F_fixed_point}) and estimating associated exponents. We end with concluding remarks in Sec.~\ref{sec:conclusions}.

\section{RGFlow architecture}
\label{sec:architecture}

Consider a lattice field theory defined on a $d$-dimensional cubic lattice. Let $\phi(\vect{x})\in \mathbb{R}$ denote a scalar field configuration sampled from the fine-grained action $\mathcal{S}_\text{UV}\left[\phi(\vect{x}); \mathbf{K}_\text{UV}\right]$, which is defined on a lattice with $(2N)^d$ sites (with $\vect{x}=(x_1,x_2,\cdots,x_d)$ coordinating the lattice site). Likewise, let $\psi(\vect{x})$ denote a configuration sampled from the coarse-grained action $\mathcal{S}_\text{IR}\left[\psi(\vect{x}); \mathbf{K}_\text{IR}\right]$, defined on a coarser lattice with $N^d$ sites. Here, $\mathbf{K}_\text{UV}$ and $\mathbf{K}_\text{IR}$ represent the sets of coupling constants at the ultraviolet (UV) and infrared (IR) scales, respectively. 

The probability density functions associated with these field configurations are given by the Boltzmann distributions (assuming unit temperature):
\begin{equation}
    \begin{split}
        P_{\text{UV}}\left[\phi\right] &= \frac{e^{-\mathcal{S}_{\text{UV}}\left[\phi;\textbf{K}_{\text{UV}}\right]}}{Z_{\text{UV}}}\,,
        \\
        P_{\text{IR}}\left[\psi\right] &= \frac{e^{-\mathcal{S}_{\text{IR}}\left[\psi;\textbf{K}_{\text{IR}}\right]}}{Z_{\text{IR}}},
    \end{split}
    \label{eqn:1}
\end{equation}
where $Z_{\text{UV}/\text{IR}}=\sum_{[\phi]/[\psi]}e^{-\mathcal{S}_{\text{UV}/\text{IR}}}$ represents the partition function. From a probabilistic point of view, a general RG transformation can be described using the conditional probability $P_\theta [\psi| \phi]$ where $\theta$ denotes the set of parameters that parametrize the local transformation (coarse-graining) rule, such that the coarse-grained (IR) and fine-grained (UV) probability density functions are related by the integral
\begin{equation}
    P_{\text{IR}}\left[\psi\right] = \int \mathcal{D}\phi \ P_\theta [\psi| \phi] P_{\text{UV}}\left[\phi\right].
    \label{eqn:RG}
\end{equation}

It is evident that there is substantial freedom in the choice of RG transformations as parameterized by some $\theta$, and most choices do not yield physically meaningful predictions. Traditionally, RG transformations are optimized by imposing constraints on the physical properties of the transformed action. For instance, one class of methods computes magnetic scaling exponents independently from both the free energy and the correlation functions, and then enforces their consistency \cite{kadanoff_numerical_1975,wilson_renormalization_1975}. Another approach involves optimizing the variational lower and upper bounds on the free energy prior to the transformation  \cite{migdal_phase_1975,migdal_recursion_1975,kadanoff_notes_1976}. However, these methods typically involve a limited number of tunable parameters, due to the computational complexity and analytical effort required. More recently, optimization principles based on information theory have been introduced. Unlike the traditional approaches, these methods directly optimize probability densities, therefore substantially reducing computational overhead. This advancement enables the RG transformation to be represented by much more complex models, such as restricted Boltzmann machines \cite{koch-janusz_mutual_2018,lenggenhager_optimal_2020}, or even neural networks \cite{hu_machine_2020-1}. 

In this study, we adopt the minimal bulk mutual information principle formulated previously in Ref.~\cite{hu_machine_2020-1}.
This principle posits that the RG transformation should eliminate irrelevant features---defined as those degrees of freedom (DOFs) that exhibit minimal mutual information. In the ideal case, this minimum is achieved when the decimated DOFs are mutually independent and behave as decoupled random noise. The motivation for this principle comes from the well-known fact that mutual information provides an upper bound on statistical correlation functions (see Appendix A). By minimizing the mutual information among the discarded DOFs, the optimized RG transformation effectively suppresses irrelevant correlations and compresses the relevant physical information into the coarse-grained variables. Furthermore, Ref.~\cite{hu_machine_2020-1} establishes the equivalence between the minimal bulk mutual information principle and the more widely studied maximal real-space mutual information principle \cite{koch-janusz_mutual_2018,lenggenhager_optimal_2020}, which aims to maximize the mutual information between the coarse-grained field and its surrounding fine-grained environment. In our formulation, we model the irrelevant features [denoted by $\xi(\textbf{x})$] as independent Gaussian random variables with distribution $P[\xi]=\mathcal{N}(0, \mathbb{1})$, so that the minimal bulk mutual information criterion is satisfied by design.

Additionally, we have chosen our RG map to be bijective. Instead of discarding the irrelevant features, as in traditional RG, we keep them. The coarse-grained configuration and the generated random noises then jointly form a ``latent configuration,'' which shares the same dimension as the fine-grained one. Such kind of information-preserving RG transformation has been successfully applied to a variety of field theory models~\cite{qi_exact_2013,lee_exact_2016,gu_holographic_2016}. Finally, we implement our bijective RGFlow network using so-called normalizing flow architecture~\cite{rezende_variational_2015,papamakarios_normalizing_2021,kobyzev_normalizing_2021}, whose network parameters become the aforementioned set of parameters $\theta$ characterizing the RG transformation. The optimality is achieved through training the neural network.

\begin{figure*}
    \centering
    \includegraphics[width=5in]{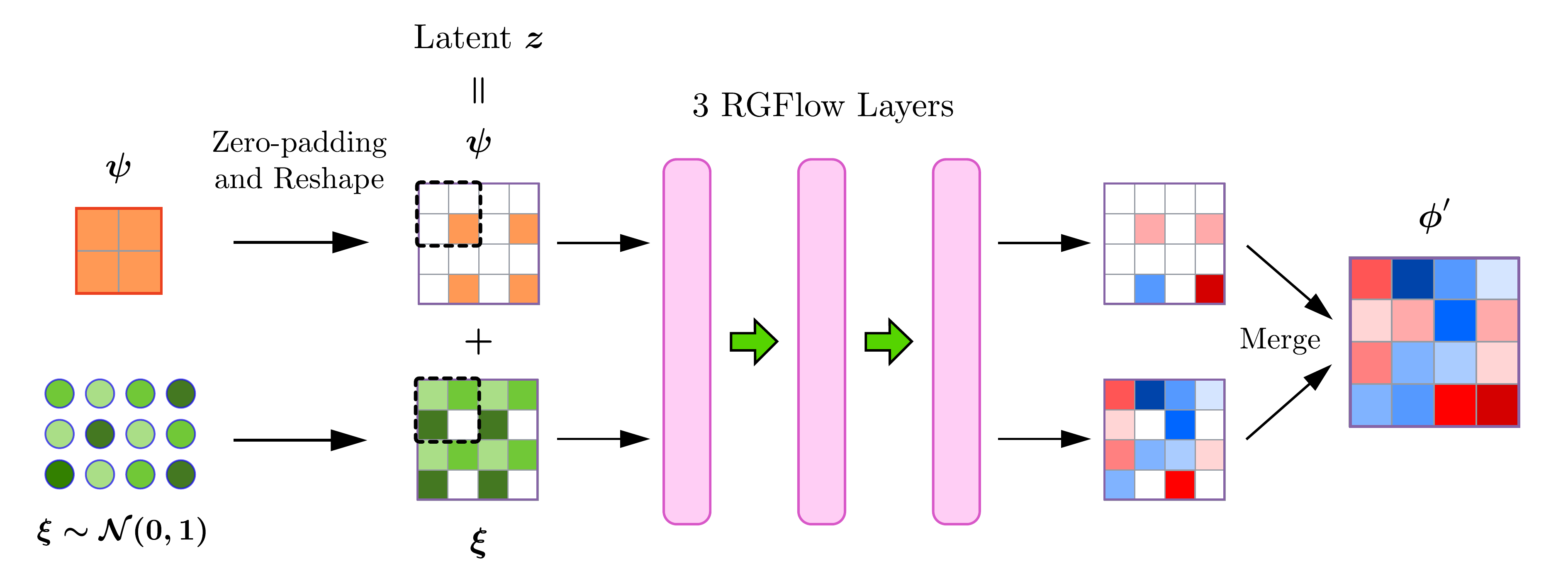}
    \caption{An example of the proposed RGFlow scheme in 2D. Coarse-grained configurations $\psi$ (in orange) and irrelevant features $\xi$ (in greens) are sampled from corresponding expressions before zero-padded and reshaped into size $4\times 4$. The expanded configurations are consisted of four identical RG cells of size $2\times 2$ (The dashed black boxes). There are one coarse-grained site and three irrelevant features in one cell. $\psi$ and $\xi$ make up the latent configuration $z$. They are then treated as two different channels and inputted to the RGFlow network, which generates the predicted fine-grained configurations $\phi'$. The colors in $\phi'$ represent field strengths.}
    \label{fig:RGFlow}
\end{figure*}

The proposed RGFlow renormalization scheme utilizes a local transformation rule in the inverse renormalization direction and generates $\phi$ from $\psi$ in the following three steps. (I) Sample a coarse-grained configuration $\psi$ and a vector $\xi$ of irrelevant features of size $(2N)^d - N^d$ from $P_{\text{IR}}\left[\psi\right]$ and $P[\xi]$ respectively. (II) $\psi$ and $\xi$ are expanded using zero-padding and reshaped into size $(2N)^d$. The resulting configuration consists of $N^d$ RG cells of $2^d$ sites. Each RG cell contains one lattice site from $\psi$ and $2^d - 1$ sites from $\xi$. (III) This configuration is given as an input to the network, which generates a predicted fine-grained configuration $\phi'(\textbf{x})$. The rescaling factor for our RG scheme is $b = a'/a = 2$ since the configuration size has doubled following this procedure. The above RG steps are depicted in Fig.~\ref{fig:RGFlow} for the case $d = 2$. It is important to notice that our RGFlow scheme is similar to the NeuralRG~\cite{li_neural_2018}. Both architectures treat the irrelevant DOFs as mutually independent Gaussian noises and use a neural network to approximate the bijective RG transformation.
Further, the RGFlow model could be viewed as the first layer in the hierarchical NeuralRG structure. However, there still exists a major difference. In the NeuralRG model, the fine-grained configuration is mapped entirely to a random noise, whereas in the RGFlow scheme, we still keep track of the coarse-grained coupling constants $\textbf{K}_{\text{IR}}$ and therefore their renormalization.

To further simplify the notation, we define $z = (\psi,\xi)$ to be our latent configuration 
and we denote the RG transformation by $z = T_\theta(\phi)$ where $\theta$ are the network parameters.
We can express the model distribution $P_{\text{UV}}'[\phi]$ for the RGFlow predicted fine-grained configurations $\phi'$ using $P_{\text{IR}}[\psi]$ and $P[\xi]$ as 
\begin{equation}
    P_{\text{UV}}'[\phi] = P[z] /
   \det J_{T^{-1}}(z)\,,
   \quad
   J_{T^{-1}}(z) =
    \frac{\delta T_\theta^{-1}(z)}{\delta z}\,.
\label{eqn:3}
\end{equation}
Here $J_{T^{-1}}(z)$ is the Jacobian matrix of the transformation and $P[z]$ is the probability distribution function of $z$:
\begin{equation}
    P[z] = P_{\text{IR}}[\psi]P[\xi].
    \label{eqn:add1}
\end{equation}
Using equation \eqref{eqn:3} and \eqref{eqn:add1} we can explicitly express the conditional probability in equation \eqref{eqn:RG} as
\begin{equation}
    P_\theta [\psi|\phi] = \frac{\det J_{T^{-1}}(z)}{P[\xi]} \delta(T_\theta^{-1}(z)-\phi).
\end{equation}

Our goal is to construct $T_\theta$ such that the generated fine-grained configurations $\phi'(\textbf{x})$ have the same probability distribution as true $\phi$. This requires maximizing the similarity between $P_{\text{UV}}'[\phi]$ and the ideal distribution $P_{\text{UV}}[\phi]$ from Eq.~\eqref{eqn:1}. A popular measure of the similarity between two probability distributions is the Kullback–Leibler (KL) divergence. However, since the partition function $Z_{\text{UV}}$ and $Z_{\text{IR}}$ are intractable for most lattice field theories, direct evaluation of the KL divergence is impossible. Therefore, we instead choose the Fisher divergence, which takes the following form:
\begin{equation}
    \begin{split}
        D_{\text{F}}\left(P_{\text{UV}}' \ || \ P_{\text{UV}}\right) &= \displaystyle \mathop{\mathbb{E}}_{\phi \sim P_{\text{UV}}'[\phi]} \left\Vert \frac{\delta \log P_{\text{UV}}'[\phi]}{\delta \phi} - \frac{\delta \log P_{\text{UV}}[\phi]}{\delta \phi}\right\Vert_{2}^{2} \\
        &= \displaystyle \mathop{\mathbb{E}}_{\phi \sim P_{\text{UV}}'[\phi]} \left\Vert \frac{\delta \mathcal{S}_{\text{UV}}'[\phi;\textbf{K}_{\text{IR}}]}{\delta \phi} - \frac{\delta \mathcal{S}_{\text{UV}}[\phi;\textbf{K}_{\text{UV}}]}{\delta \phi}\right\Vert_{2}^{2}.
    \end{split}
    \label{eqn:4}
\end{equation}
Equation~\eqref{eqn:4} can be easily evaluated since we have from Eq.~\eqref{eqn:3} and~\eqref{eqn:add1}
\begin{align}
    \mathcal{S}_{\text{UV}}'[\phi;\textbf{K}_{\text{IR}}] &=\mathcal{S}_{\text{z}}[z;\textbf{K}_{\text{IR}}] + \log \det J_{T^{-1}}(z)\,,
\label{eqn:5}\\
\mathcal{S}_{\text{z}}[z;\textbf{K}_{\text{IR}}] &= \mathcal{S}_{\text{IR}}[\psi;\textbf{K}_{\text{IR}}] + \frac{1}{2}\xi^T\xi\,,
\end{align}
where $\mathcal{S}_{\text{z}}[z;\textbf{K}_{\text{IR}}]$ is the latent action.
The Jacobian matrix of the RG transformation is $J_T(\phi) = \frac{\delta z}{\delta \phi} = \frac{\delta T_\theta (\phi)}{\delta \phi}$. 
Applying the chain rule for derivatives, we transform the Fisher divergence to
\begin{equation}
        D_{\text{F}}\left(P_{\text{UV}}' \ || \ P_{\text{UV}}\right) =
        \displaystyle \mathop{\mathbb{E}}_{z \sim P_{\text{z}}[z]} \left\Vert \left[\frac{\delta (\mathcal{S}_{\text{z}}[z;\textbf{K}_{\text{IR}}] + \log \det J_{T^{-1}}(z))}{\delta z} - \frac{\delta \mathcal{S}_{\text{UV}}[T_\theta^{-1}(z);\textbf{K}_{\text{UV}}]}{\delta z}\right]J_T(\phi)\right\Vert_{2}^{2}.
\label{eqn:6}
\end{equation}
Since for a meaningful RG map $J_T(\phi)$ should not vanish, we drop this factor and define our training loss function to be
\begin{equation}
    L(\textbf{K}_{\text{IR}}, \theta) = \displaystyle \mathop{\mathbb{E}}_{z \sim P_{\text{z}}[z]} \left\Vert \left[\frac{\delta (\mathcal{S}_{\text{z}}[z;\textbf{K}_{\text{IR}}] + \log \det J_{T^{-1}}(z))}{\delta z} - \frac{\delta \mathcal{S}_{\text{UV}}[T_\theta^{-1}(z);\textbf{K}_{\text{UV}}]}{\delta z}\right]\right\Vert_{2}^{2}.
    \label{eqn:7}
\end{equation}
This loss function has the same global minimum (zero) as $D_F$. 
Evaluating $L(\textbf{K}_{\text{IR}}, \theta)$ requires computing the Jacobian $\det J_{T^{-1}}(z)$. To make it computationally efficient, we implement the mapping $T_{\theta}$ using several RealNVP modules~\cite{dinh_density_2017}. The corresponding Jacobian matrix is triangular, and so its determinant is equal to the product of the diagonal elements of $J_{T^{-1}}(z)$. To utilize the RealNVP module, we treat the zero-padded $\psi$ and $\xi$ as two input channels. After passing each layer, the order of the two channels switch. Therefore, two consecutive RealNVP modules make up one RGFlow layer, which functions as a basic building block of the network. As demonstrated in the 2D example in Fig.~1, the upper channel contains only the coarse-grained fields $\psi$ while the lower channel holds the remaining irrelevant features. After being transformed by the neural network, the two channels are merged to reconstruct the predicted $\phi'$.

During training, a batch of configurations $(\psi, \xi)$ are sampled from $P_{\text{IR}}\left[\psi\right]$ and $P[\xi]$ to calculate $L(\textbf{K}_{\text{IR}}, \theta)$ at each step. A stochastic optimizer is then applied to compute the gradients and optimize $T_\theta^{-1}$ and $\textbf{K}_{\text{IR}}$ simultaneously. After reaching a local optimum, a unique set of predicted coarse-grained coupling $\textbf{K}_{\text{IR}}$ will be associated with the fined-grained couplings $\textbf{K}_{\text{UV}}$, which are kept fixed during training. By separately probing different sets of $\textbf{K}_{\text{UV}}$, we can study the RG flow at any location in the parameter space. In other words, we do not impose any limitations on the coupling constants or space dimension (unlike, e.g., the $\epsilon$-expansion~\cite{WILSON197475,RevModPhys.46.597}). Moreover, since a unique neural network is trained for every pair of $\textbf{K}_{\text{UV}}$ and $\textbf{K}_{\text{IR}}$, the learned RG map is automatically adjusted to extract the most relevant feature. Therefore, we expect that the proposed method may be suitable for studying the RG flows of systems with multiple phases. 

\section{A solvable one-dimensional model}
\label{sec:analytical}

The described RGFlow scheme can be solved exactly for the case of a one-dimensional (1D) Gaussian model with the action
\begin{equation}
    \mathcal{S}_{\text{UV}}[\phi;r_{\text{UV}}] = \frac{1}{2}
    \sum_{j = 0}^{2 N - 1} \left[ \left(\phi_{j + 1} - \phi_j \right)^2 + r_{\text{UV}} \phi_j^2\right]\,,
\label{eqn:gaussian_action}
\end{equation}
where $\phi_j$ is a scalar field on $j$th site
of an $2 N$-site lattice with the periodic boundary conditions and a unit lattice spacing constant.
It it convenient to double the unit cell to match the latent model, see below, and introduce the two-component fields
$\Phi_k = (\phi_{2 k}, \phi_{2k + 1})^T$, where $k = 0, 1, \ldots, N - 1$.
We define the Fourier transform $\tilde\Phi(q) = N^{-1 / 2} \sum_{k = 0}^{N - 1} e^{-i q k} \Phi_k$ of these fields and rewrite the model action as follows:
\begin{equation}
    \mathcal{S}_{\text{UV}} = \frac{1}{2}
    \sum_{n = 0}^{N - 1} \tilde\Phi(q_n)^\dagger \Omega_{\text{UV}}(q_n)
    \tilde\Phi(q_n)\,,
    \quad
    \Omega_{\text{UV}}(q_n) = 
\begin{pmatrix}
    r_{\text{UV}} + 2  & -(1 + e^{-iq_n})\\
    -(1 + e^{iq_n})        & r_{\text{UV}} + 2
\end{pmatrix},
\quad
    q_n = \frac{2 \pi n}{N}\,.
    \label{eqn:S_UV}
\end{equation}
Note that the eigenvalues $\omega_\pm(q) = r_{\text{UV}} + 2 \pm 2 \cos q$
of the matrix $\Omega_{\text{UV}}(q)$ represent the band dispersion in the problem.

In turn, the action of the latent model is
\begin{equation}
    \begin{split}
        \mathcal{S}_{\text{Z}} &= \frac{1}{2} \sum_{j = 0}^{N - 1}
        \left\{
        \left[ \left(\psi_{j + 1} - \psi_j \right)^2 + r_{\text{IR}} \psi_j^2\right] + \xi_j^2 \right\}.
    \end{split}
    \label{eqn:latent_action}
\end{equation}
We again introduce two-components fields $Z_j = (\psi_j, \xi_j)^T$ and
the corresponding Fourier transforms $\tilde{Z}(q_n)$,
in terms of which the latent action becomes
\begin{equation}
    \mathcal{S}_{\text{Z}} = \frac{1}{2}
    \sum_{n = 0}^{N - 1} \tilde{Z}(q_n)^\dagger \Omega_{\text{Z}}(q_n)
    \tilde{Z}(q_n)\,,
    \qquad
\Omega_{\text{Z}}(q_n) =
\begin{pmatrix}
    r_{\text{IR}} + 2 - 2 \cos q_n & 0\\
                        0          & 1
\end{pmatrix}\,.
\label{eqn:S_IR}
\end{equation}
Since both $\mathcal{S}_{\text{UV}}$ and $\mathcal{S}_{\text{Z}}$ are quadratic,
the exact global minimum $L = 0$ of the loss function $L$ from Eq.~\eqref{eqn:7} can be achieved by a linear RG transformation $T_\theta$
of a convolution type such that
\begin{equation}
\tilde{Z}(q) = \tilde{T}_\theta(q) \tilde{\Phi}(q)\,,
\qquad
(\tilde{T}_\theta)^\dagger \Omega_{\text{Z}} \tilde{T}_\theta =  \Omega_{\text{UV}}\,.
\label{eqn:S_z_from_S_UV}
\end{equation}
Here, as above, the tilde denotes the Fourier transform, which is defined as follows.
It suffices to consider $T_\theta$ with the following general size-$3$ convolution kernels
for $\psi$'s and $\xi$'s:
\begin{equation}
\mathcal{C}_\psi = \{a, c, d \}\,,
\qquad
\mathcal{C}_\xi = \{f, g, h \}
\label{eqn:C_1_C_2}
\end{equation}
with real $a, c, d, f, g$ and $h$. Matrix $\tilde{T}_\theta$ is then given by
\begin{equation}
\tilde{T}_\theta(q_n) =
\begin{pmatrix}
    a+ e^{i q_n} d      & c\\
  e^{i q_n} g  &  f + e^{i q_n} h
\end{pmatrix}\,.
\label{eqn:tilde_T_theta}
\end{equation}
Equation~\eqref{eqn:S_z_from_S_UV} has the exact solution
$a = d = 0$, $b = \pm {1} / {\sqrt{r_{\text{UV}} + 2}}$,  $c = f = h = -1 / g$, and $g = \pm \sqrt{r_{\text{UV}} + 2}$.
The corresponding renormalized coupling is
\begin{equation}
    r_{\text{IR}} = 4 r_{\text{UV}} + r_{\text{UV}}^2.
    \label{eqn:scaling}
\end{equation}
The first term on the right-hand side of this equation agrees with the scaling law $r_{\text{IR}} = b^2 r_{\text{UV}}$ known to apply in the continuum limit. The second term can be considered a lattice correction.
Importantly, this lossless RG scheme preserves the behavior of the correlation functions, such as the exponential decay of the two-point function
\begin{equation}
    \langle \phi_m \phi_n \rangle
    = \int\limits_{-\pi}^{\pi} \frac{d q}{2\pi} \frac{e^{i (m - n) q}}{r_{\text{UV}} + 2 - 2 \cos q}
    = \frac{1}{2 \sinh \varkappa_{\text{UV}}}\, e^{-\varkappa_{\text{UV}} |m - n|}\,.
\label{eqn:Gamma}
\end{equation}
(This formula applies for the infinite lattice, $N \to \infty$.) Parameter
\begin{equation}
    \varkappa_{\text{UV}} = \arccosh \left(1 + \frac{1}{2} r_{\text{UV}}\right)
\label{eqn:l}
\end{equation}
is the inverse correlation length. Equation Eq.~\eqref{eqn:scaling} entails
$\varkappa_{\text{IR}} = 2 \varkappa_{\text{UV}}$, so that the two-point functions before and after the RG transformation are related by
\begin{equation}
    \langle \psi_0 \psi_{i} \rangle
    = Z \langle \phi_0 \phi_{2i} \rangle\,,
\label{eqn:psi_psi}
\end{equation}
as required for the case $b = 2$ we are dealing with. This relation can be also seen directly from the form of $\mathcal{C}_{\psi}$ kernel; in particular, the coefficient of proportionality in Eq.~\eqref{eqn:psi_psi}, i.e., the field renormalization factor is simply $Z = c^2 = (r_{\text{UV}} + 2)^{-1}$.

The above results are also in agreement with those derived from the standard real-space RG procedure employing decimation. In such an approach, the Gaussian chain is partitioned into unit cells each containing one even-indexed site and one odd-indexed site, $\Phi_j = (\phi_{e,j}, \phi_{o,j})$, where $j=0,1, \ldots, N-1$. Under this representation, the Gaussian action \eqref{eqn:gaussian_action} can be expressed as follows:

\begin{equation}
    \mathcal{S}_{\text{UV}}[\phi;r_{\text{UV}}] = \sum_{j=0}^{N-1}\mathcal{S}_{j}
    = \sum_{j=0}^{N-1}  \frac{1}{2}(\phi_{o,j} - \phi_{e,j})^2 + \frac{1}{2}(\phi_{e,j+1} - \phi_{o,j})^2 + \frac{r_{\text{UV}} }{2} (\frac{1}{2}\phi_{e,j}^2 + \frac{1}{2}\phi_{e,j+1}^2 + \phi_{o,j}^2).
\label{eqn:gaussian_action_new}
\end{equation}
After completing the square with respect to $\phi_{o,j}$, we can separate $\mathcal{S}_{j}$ into two parts
\begin{equation}
\begin{split}
    \mathcal{S}_{j} &= \mathcal{S}_{\text{N}, j}[\xi] + \mathcal{S}_{\text{IR}, j}[\psi;r_{\text{IR}}]. \\
    &= \frac{1}{2}\xi_j^2 + \frac{1}{2}\left[(\psi_{j+1} - \psi_j)^2 + \frac{1}{2} r_{\text{IR}}(\psi_j^2 + \psi_{j+1}^2)\right],
\end{split}
    \label{eqn:complete_square}
\end{equation}
with the following definition 
\begin{equation}
    \begin{split}
        \xi_j &= -\frac{1}{\sqrt{2 + r_{\text{UV}}}}\phi_{e,j} + \sqrt{2 + r_{\text{UV}}}\phi_{o,j} - \frac{1}{\sqrt{2 + r_{\text{UV}}}}\phi_{e,j+1}, \\
        \psi_j &= \frac{1}{\sqrt{2 + r_{\text{UV}}}}\phi_{e,j}, \ \ \ \psi_{j+1} = \frac{1}{\sqrt{2 + r_{\text{UV}}}}\phi_{e,j+1}, \\
        r_{\text{IR}} &= 4r_{\text{UV}} + r_{\text{UV}}^2.
    \end{split}
    \label{eqn:rg_def}
\end{equation}
It is evident that $\mathcal{S}_{N, j}[\xi]$ corresponds to unit Gaussian noise and $\mathcal{S}_{\text{IR}, j}[\psi;r_{\text{IR}}]$ is consistent with the Gaussian action \eqref{eqn:gaussian_action} under periodic boundary conditions. In addition, the definitions provided in \eqref{eqn:rg_def} reproduce both the kernel parameters and the scaling relations obtained by minimizing the loss function \eqref{eqn:S_z_from_S_UV}. This correspondence allows for a natural interpretation of the kernel $\mathcal{C}_\psi$ as performing a decimation operation, effectively selecting every second site in the fine-grained configuration. Simultaneously, the kernel $\mathcal{C}_\xi$ captures the discarded DOFs and encodes them as Gaussian noise. In conventional real-space RG, the transformation is completed by integrating out the Gaussian noise $\mathcal{S}_{N, j}[\xi]$, thereby making the procedure non-invertible. In contrast, the RGFlow framework retains this information, thus preserving bijectivity and enabling a fully reversible transformation.

\section{Numerical results}
\label{sec:results}

\subsection{Model}
\label{sub:model}

We now focus on the 2D $\phi^4$ theory on a square lattice with the periodic boundary conditions and action
\begin{equation}
    \mathcal{S}[\phi;r,u] = \frac12 \sum_{\langle i j \rangle}
    \left(\phi_{i} - \phi_j \right)^2
    + \sum_i \frac{r}{2}\, \phi_i^2 + \frac{u}{4}\, \phi_i^4\,,
    \label{eqn:8}
\end{equation}
where $\phi_i$ is a real valued scalar field defined on lattice site $i$ and $\langle i j \rangle$ is the bond between nearest-neighbor lattice sites $i$, $j$.
The coupling constants $\textbf{K} = (u, r)$ are dimensionless. Constant $u$ must be non-negative for stability.

At large $u$, the lattice $\phi^4$ theory can be mapped onto the Ising model $S = -J \sum_{\langle i j \rangle} \sigma_i \sigma_j$ with the dimensionless exchange constant $J = r / u$ and discrete site variables $\sigma_i = \pm 1$.
From Onsager's exact solution of the 2D Ising model
we know that the system undergoes the ferromagnetic transition at
\begin{equation}
        r \simeq -c_1 u\,,
        \quad c_1 = \frac12 \ln \left(1 + \sqrt2\,\right),
        \quad u \gg 1.
    \label{eqn:critical_large}
\end{equation}
It is believed that at smaller $u$ the transition remains within the Ising universality class. The corresponding critical line in the $(u,r)$ parameter space has been studied by Monte-Carlo simulations~\cite{loinaz_monte_1998,sugihara_density_2004,De_investigations_2005,schaich_improved_2009,milsted_matrix_2013} and other methods (\cite{Rychkov_hamiltonian_2015, serone_4_2018, Delcamp_computing_2020}, etc).
For small $u$, these calculations predict the critical $r$ to be
\begin{equation}
        r \simeq \frac{u}{c_0} + \frac{3}{4\pi}\, u \ln\left(\frac{u}{32 c_0}\right),
        \quad 0 < u \ll 1\,.
    \label{eqn:critical_small}
\end{equation}
The second term in this equation comes from the one-loop renormalization of the mass term $r$ on the square lattice.
At $u \to 0$, most recent calculations predict $c_0 \approx 11$, see Table~I in Ref.~\cite{Delcamp_computing_2020}.
However, using $c_0 = 9.9$ gives an excellent overall fit to the Monte-Carlo data~\cite{loinaz_monte_1998, De_investigations_2005} within the range $0 < u < 2$, see Fig.~\ref{fig:RG}.

It is easy to see that the phase-transition line approximated by Eq.~\eqref{eqn:critical_small} must be a streamline of the RG flow, in the following sense.
According to the modern Wilsonian formulation of the RG, the exact RG flow involves infinite number of coupling constants besides $r$ and $u$, e.g., those for $\phi^6$, $\phi^8$, \textit{etc}. terms in the action.
On the $(u, r)$-plane all those additional coupling constants are equal to zero.
Importantly, any streamline of the exact RG flow that starts from a point on the phase transition line in the $(u, r)$-plane eventually arrives at the same fixed point in the higher-dimensional space. The union of these streamlines defines the RG critical surface~\cite{Shankar_2017}. The projection of the flow vectors of this critical surface onto the $(u, r)$ plane must be tangent to the phase-transition line.
This reasoning holds as long as the coupling constants evolve continuously, which happens when the rescaling factor $b$ approaches unity. In this work, we adopt
$b=2$, a common choice in real-space RG. Therefore, the RGFlow-computed critical line may somewhat deviate from the exact phase-transition line.

We expect that the critical line contains two fixed points,
the Gaussian fixed point $u = r = 0$ and a nonperturbative (Wilson-Fisher-like) fixed point at $u = u^* > 0$, $r = r^* < 0$.
The latter is the approximation of the projection of the true multi-dimensional fixed point onto the $(u,r)$ plane. The RG flow near $(u^*, r^*)$ can be used to estimate the correlation length critical exponent $\nu$~\cite{Shankar_2017}.
Our goal is to compute the critical line by the RGFlow scheme to verify these expectations and to investigate any additional structures of the flow in a larger parameter space around the critical line.

\subsection{Critical line and critical exponents}
\label{sub:critical}

To construct an effective RG map, we build a RealNVP module using 2D convolution layers with the same kernel size of $3\times 3$. In total, three RGFlow layers are included in the final neural network. We demonstrate the entire RGFlow network structure in Fig.~\ref{fig:RGFlow}. Inside this architecture, the convolution layers collect local correlations. This local information is later shared at different scales when passing through the RGFlow layers. As a quick demonstration of the proposed RG scheme, we choose a minimal configuration size of $2\times 2$ for $\psi$ sampled from $\mathcal{S}_{\text{IR}}$.
During training, a batch of 100 $\psi$ and $\xi$ configurations are sampled at each step. The Adam optimizer~\cite{Kingma_adam_15} with an initial step size of $h = 10^{-3}$ is applied to minimize the loss function (\ref{eqn:7}). After every 5000 iterations, $h$ is reduced by 85\%. A total of 40,000 training iterations are scheduled for each session.

\begin{figure}
    \centering
    \includegraphics[width=5in]{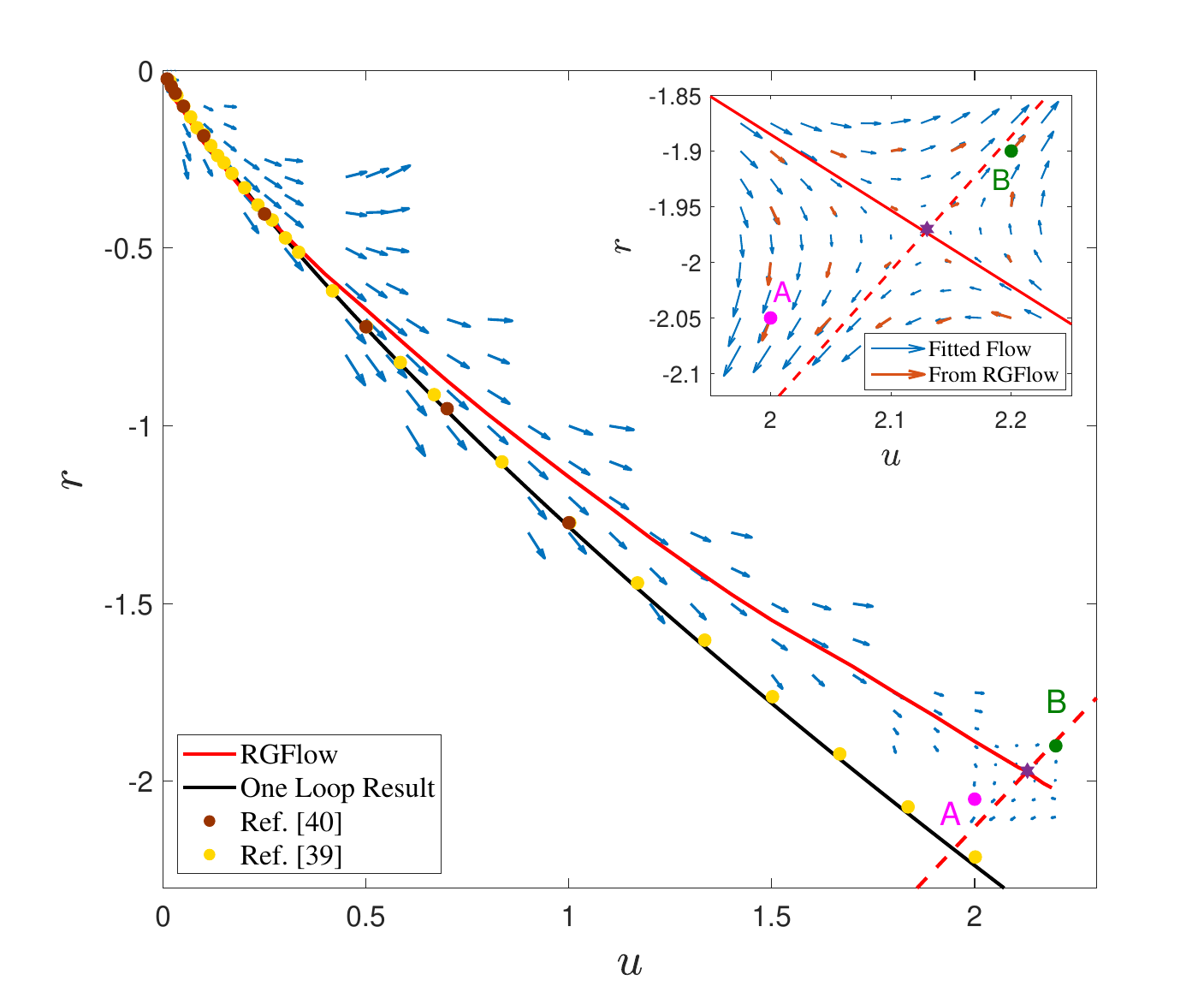}
    \caption{The predicted RG flow diagram for the 2D $\phi^4$ model using the RGFlow algorihthm. 
    The blue arrows are vectors $\Delta \textbf{K} = \textbf{K}_{\text{IR}}-\textbf{K}_{\text{UV}}$. The purple star pinpoints the fixed point of the saddle-point structure. The red solid line represents the critical line predicted by RGFlow. 
    As a comparison, we graph equation \eqref{eqn:critical_small} in solid black and present the Monte Carlo results reported in~\cite{schaich_improved_2009} and \cite{De_investigations_2005} as brown and yellow dots respectively. The corresponding correlation maps at points A (pink) and B (green) are provided in Fig.~\ref{fig:maps}. The red dashed line indicates the relevant direction of the flow. The inset figure provides a detailed view of the RG flow around the saddle point. Here, the red vector field represents the RGFlow predicted flow while the blue vector field is the result from equation \eqref{eqn:11}. Other features remain the same as the outer graph.}
    \label{fig:RG}
\end{figure}

We applied the RGFlow algorithm to a number of points in the $(u, r)$ space near the expected position of the critical line. The results are displayed as a vector field $\Delta \textbf{K} = \textbf{K}_{\text{IR}}-\textbf{K}_{\text{UV}}$ in Fig.~\ref{fig:RG}. We observe three types of parameter flows: the first one goes to $r=-\infty$ (indicating the symmetric phase), the second one goes to $r = -\infty$ (the broken symmetry phase), the remaining third one flows into the saddle point marked by the purple star. We find the location of this fixed point $\textbf{K}^*=( u^*, r^*)$ as follows. We first apply the Least Squares method \cite{coleman_convergence_1994,coleman_interior_1996} to fit the saddle point structure to a linear vector field
\begin{equation}
    \Delta \mathbf{K}(\mathbf{K}_{\text{UV}}) = \vect{M} \mathbf{K}_{\text{UV}} + \vect{d},
    \label{eqn:11}
\end{equation}
where $\vect{M}$ and $\vect{d}$ represent a $2\times 2$ matrix and a $2\times 1$ vector respectively. Next, we compute $\mathbf{K}^* = -\vect{M}^{-1} \vect{d}$ such that $\Delta \mathbf{K}(\mathbf{K}^*) = 0$. In this numerical study, we find that the fixed point is located at $r^* = -1.97\pm 0.12$ and $u^* = 2.13\pm 0.13$. 

The critical line is defined to be the streamline of the vector field $\Delta \mathbf{K}$ that connects the Gaussian fixed point at the origin and the Wilson-Fisher-like fixed point $\textbf{K}^*$. The result is displayed as the solid red curve in Fig.~\ref{fig:RG},
which was computed using the \texttt{streamline} function of MATLAB.
Fitting the RGFlow critical line to Eq.~\eqref{eqn:critical_small} at $0 < u < 2$, we obtain $c_0 = 6.44 \pm 0.06$,
which is somewhat smaller than $c_0 = 9.9$ from the Monte-Carlo data (the black curve). Note,
however, that this result was obtained from a tiny sample size of $2\times 2$. Presumably, it would become more accurate if larger sample sizes are used. 

The critical exponents are estimated by linearizing the RG equation at the fixed point:
\begin{equation}
    \begin{split}
        \textbf{K}_{\text{IR}} - \textbf{K}^{*} &= \frac{\partial \textbf{K}_{\text{IR}}}{\partial \textbf{K}_{\text{UV}}} \Bigg\rvert_{\textbf{K}^{*}} (\textbf{K}_{\text{UV}} - \textbf{K}^{*}) \\
        &= (\vect{M}+\mathbb{1}) (\textbf{K}_{\text{UV}} - \textbf{K}^{*}).
    \end{split}
    \label{eqn:12}
\end{equation}
The Jacobian matrix $\vect{M} + \mathbb{1}$ of this linearized flow has two eigenvalues, $\lambda_t$ and $\lambda_h$. Following the convention~\cite{cardy_scaling_1996}, we define $|\lambda_t| > 1$ as the relevant eigenvalue and its corresponding eigenvector to be the relevant direction. In turn, $|\lambda_h| < 1$ is the irrelevant eigenvalue and its corresponding eigenvector is the irrelevant direction. As demonstrated in Fig.~\ref{fig:RG}, the relevant direction is along with the critical line while the irrelevant direction is depicted as the dashed red line.

The critical exponent $\nu$ is calculated using
\begin{equation}
    \nu = \frac{\log b}{\log |\lambda_t|}\,,
    \label{eqn:13}
\end{equation}
where $b = 2$. 
We obtain $\nu = 0.885\pm 0.015$, which differs from the exact value of $\nu = 1$ by about $10\%$.
The main cause of this discrepancy is presumably the extremely small sample size ($2 \times 2$).
Another possible reason is neglecting higher-order interactions ($\phi^6$, \textit{etc}.) mentioned above. A similar situation has been reported in a recent study of a ML-based RG for the 2D Ising model co-authored by one of us~\cite{hou_machine_2023}. By allowing for additional coupling terms beyond nearest neighbor in the predicted UV Hamiltonian, an improved prediction of the critical coupling constant for the Ising model was observed. We envision that a better estimation of $\nu$ can be achieved by similarly extending our UV action~\eqref{eqn:8}.

\begin{figure}
    \centering
    \includegraphics[width=5in]{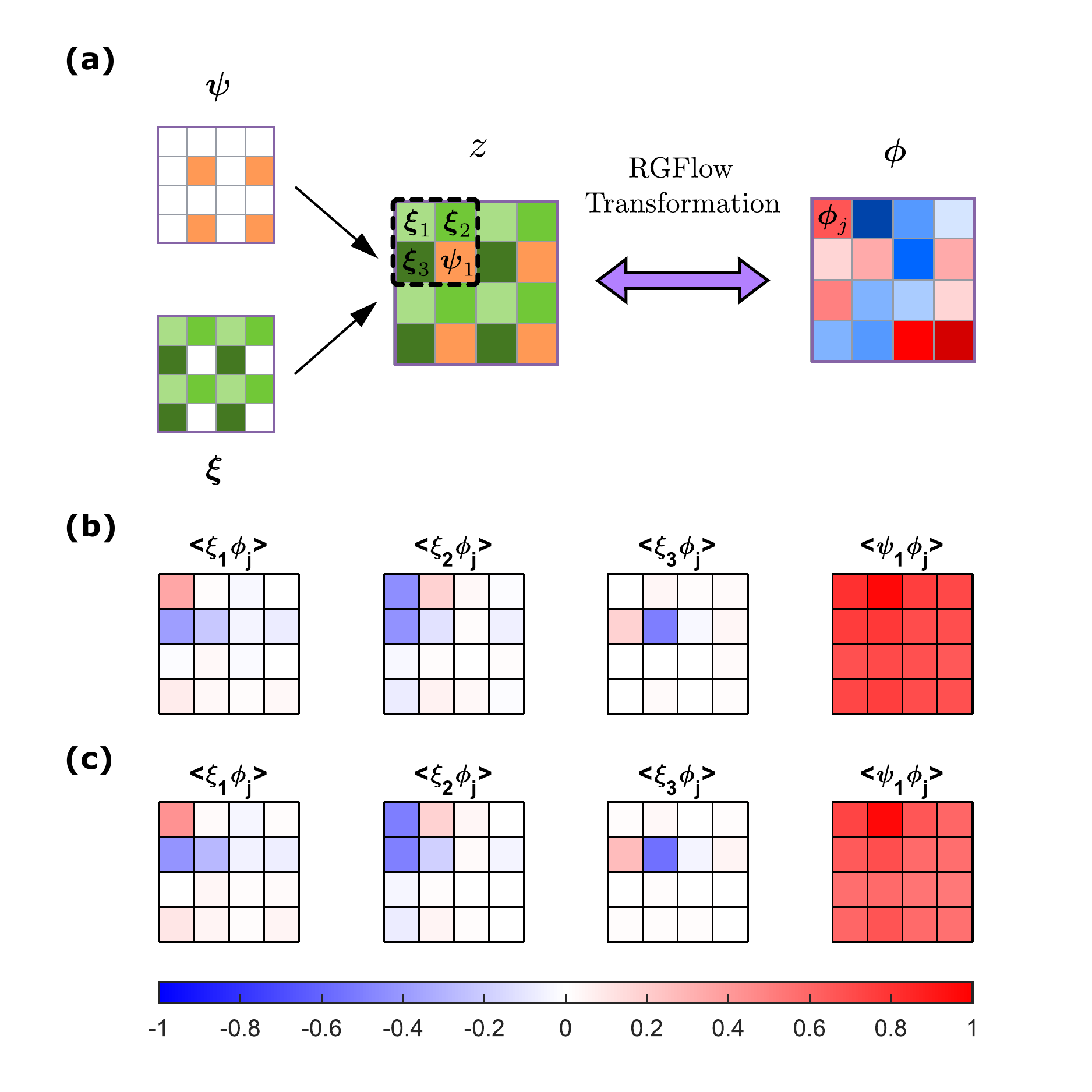}
    \caption{Correlation maps between sites in the RG unit cell and the fine-grained configuration. In panel (a), the unit cell of interest is highlighted by the dashed box. The correlation maps (\ref{eqn:14}) between four cell sites $\xi_1$, $\xi_2$, $\xi_3$, and $\psi_1$ and all fine-grained sites $\phi_j$ are computed at two specific locations, labeled A and B, in the flow diagram (see Fig. \ref{fig:RG}). The corresponding results are demonstrated in panels (b) and (c), respectively.}
    \label{fig:maps}
\end{figure}

To reveal the RG rules learned by the neural network we calculate the normalized correlation functions
\begin{equation}
\begin{split}
    \langle \xi_i \phi_j \rangle &= \displaystyle \mathop{\mathbb{E}}_{\phi \sim P_{\text{UV}}\left[\phi\right]} \frac{\xi_i \phi_j}{\sigma(\xi_i)\sigma(\phi_j)}\,,\\
    \langle \psi_i \phi_j \rangle &= \displaystyle \mathop{\mathbb{E}}_{\phi \sim P_{\text{UV}}\left[\phi\right]} \frac{\psi_i \phi_j}{\sigma(\psi_i)\sigma(\phi_j)}
\end{split}
\label{eqn:14}
\end{equation}
between a site $i$ in the RG transformed configuration and all sites $j$ in the fine-grained configuration $\phi$ (see Fig. \ref{fig:maps}(a). Here $\sigma(x)$ stands for the standard deviation of $x$. For each variable $\xi_i$ or $\psi_i$, the resulting correlation map is of size $4\times 4$ and can be interpreted as a generalized convolutional kernel and, equivalently, a RG kernel. Since the transformed configurations are composed of repeated $2\times 2$ RG cells (see Fig. \ref{fig:RGFlow}), it is sufficient to evaluate the correlation maps for the four constituent sites within the top-left unit cell. To capture the phase dependence of the learned RG rules, we calculate the correlation maps at two representative points, labeled $A$ and $B$, corresponding to the symmetric and symmetry-broken phases, respectively (see Fig. \ref{fig:RG}). The results are shown in Fig. \ref{fig:maps}(b) and \ref{fig:maps}(c). 

In both phases, the correlation strengths between the coarse-grained site $\psi_i$ and neighboring fine-grained fields $\phi_j$ remain roughly the same. This indicates that the RGFlow network has successfully identified a local averaging rule consistent with conventional real-space RG, which extracts long-range information. Notably, the strength of these correlations remains invariant across the phase transition, as expected for the $\phi^4$ theory.
In contrast, the correlation maps between the irrelevant features $\xi$ and the surrounding $\phi_j$ exhibit patterns with opposite signs. These results suggest that the network has learned effective finite-difference-like kernels that extract short-range interactions. Importantly, the noise correlation maps associated with $\xi$ remain phase independent and are consistent with the assumption that these irrelevant features follow a Gaussian distribution and are uncorrelated with the macroscopic phase behavior.

\section{Conclusions}
\label{sec:conclusions}

In this study, we introduced RGFlow, a neural network–based framework for learning real-space RG transformations and parameter flows for lattice field theories. Unlike traditional RG approaches, RGFlow preserves information by employing a bijective transformation that maps fine-grained configurations to coarse-grained fields and irrelevant features. During training, we utilize the Fisher divergence as a loss function to circumvent the intractability of partition functions. We analytically validated the RGFlow algorithm using the one-dimensional Gaussian model, demonstrating that it reproduces the correct parameter flow. Numerically, we applied RGFlow to the 2D $\phi^4$ model. Remarkably, even when trained on the smallest possible $2\times2$ lattice configurations, the method still identified a saddle point structure in the parameter flow diagram and predicted the critical line and the associated critical exponent with only a $\sim 10\%$ error.

The RGFlow method demonstrates several advantages. First, it automates the design and optimization of RG transformations across a wide parameter space. Second, unlike existing neural-network-based RG methods for lattice field theories~\cite{koch-janusz_mutual_2018,li_neural_2018,lenggenhager_optimal_2020,hu_machine_2020-1,chung_neural_2021,ron_monte_2021,giataganas_neural_2022,bachtis_inverse_2022,sheshmani_categorical_2023}, RGFlow implementation does not require explicit knowledge of the physics of the model. This feature broadens its applicability to more complex field theories. Third, since the training criterion of RGFlow relies only on evaluating the actions, it is suitable for strongly coupled regimes, where traditional methods may struggle.

We can think of several directions for improvement and future investigations. Currently, RGFlow generates RG flows in a point-wise manner, which can be computationally intensive for complex theories. Inspired by the RG Monotone method~\cite{hou_machine_2023}, a promising enhancement would be to introduce a second neural network to extrapolate the parameter flow for a wide range of parameter space. On the architectural side, replacing RealNVP layers with neural ordinary differential equation models~\cite{chen_neural_2019,grathwohl_ffjord_2018} could improve the approximation power. Additionally, the incorporation of the symmetry-respecting equivariance neural networks~\cite{cohen_group_2016,kondor_generalization_2018,weiler_3d_2018,cohen_general_2020} could further enhance the physical consistency of the learned RG transformations. With the above improvements, we envision that the RGFlow method would be a helpful tool to compute the RG transformation in high-dimensional and strongly interacting field theories. 

\begin{figure}
    \centering
    \includegraphics[width=\columnwidth]{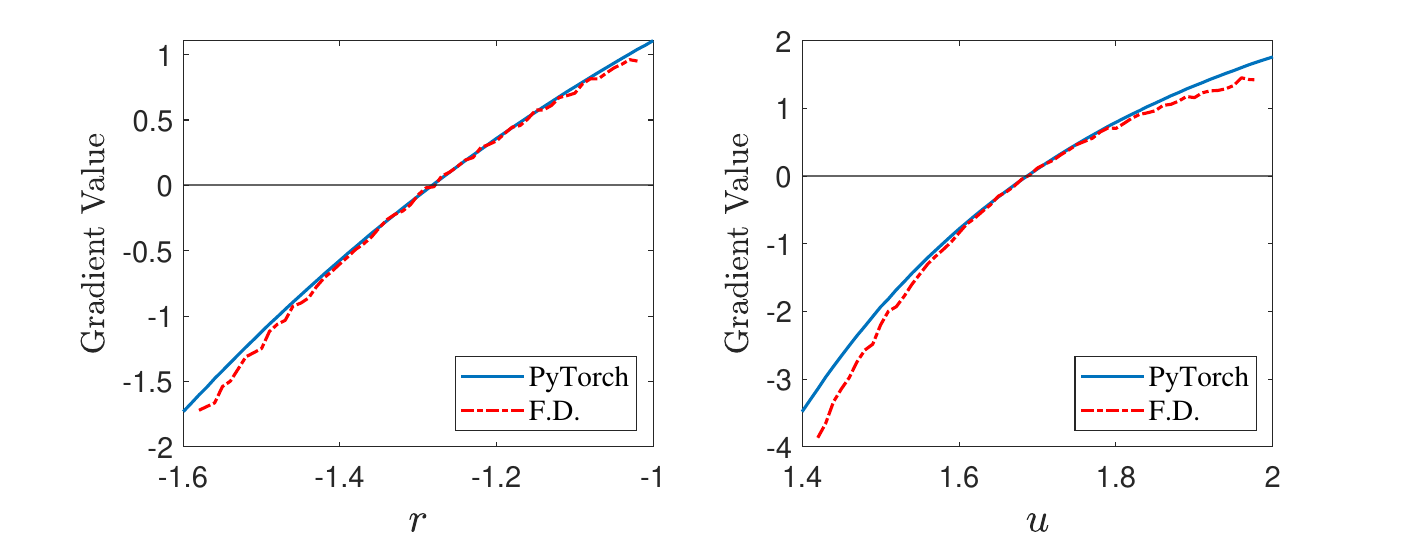}
    \caption{Comparisons between the gradients computed by PyTorch and the ideal gradients obtained via finite difference (F.D.) at one training step. The left and right panels illustrate the gradient values corresponding to the parameters $r$ and $u$, respectively, within the 2D $\phi^4$ model. The gradients produced by PyTorch exhibit excellent agreement with the finite-difference estimates, and both approaches lead to convergence toward the same local minimum. This consistency is maintained throughout the entire training process.}
    \label{fig:gradients}
\end{figure}

\emph{Note added.} By the time this paper was finalized, we noticed a small error in the gradient back-propagation process. This error was caused by the computation of the loss function \eqref{eqn:7}. During actual training, the expectation over $P_z[z]$ was approximated by the mean of \eqref{eqn:7} over configurations sampled from the distribution. However, since the samples were detached from the distribution, the gradients of $r$ and $u$ could not be passed correctly to themselves. This could introduce instability in training and lead to false local minima. Fortunately, our 2D numerical results are not substantially affected. As demonstrated in Fig. \ref{fig:gradients}, both the PyTorch gradient and the correct gradient share the same local minimum and have highly similar values. However, the behavior of PyTorch gradients for models in other dimensions remains undetermined. 

The gradient problem can be fixed by the so-called policy gradient trick, which multiplies the term inside the expectation by $1-\log P_z[z] + \log P_z[z]$. The loss function \eqref{eqn:7} then becomes
\begin{equation}
    L(\textbf{K}_{\text{IR}}, \theta) = \frac{1}{N}\sum_{z \sim P_{\text{z}}[z]} \left\Vert \left[\frac{\delta (\mathcal{S}_{\text{z}}[z;\textbf{K}_{\text{IR}}] + \log \det J_{T^{-1}}(z))}{\delta z} - \frac{\delta \mathcal{S}_{\text{UV}}[T_\theta^{-1}(z);\textbf{K}_{\text{UV}}]}{\delta z}\right]\right\Vert_{2}^{2}(1 - \log P_z[z] + \log P_z[z]).
    \label{eqn:fixed_loss}
\end{equation}
Clearly, the loss values remain unchanged. During back-propagation, the first $-\log P_z[z]$ in the parentheses is eliminated and does not pass gradients. It is then straightforward to verify that the new loss \eqref{eqn:fixed_loss} provides the correct gradient even if the same approximation is applied to the expectation. However, since calculating $\log P_z[z]$ involves evaluating the $\phi^4$ partition function, this new training method faces a significantly increased computational overhead and is beyond the scope of this work. We therefore would like to explore its effect in the follow-up studies. 

\appendix

\section{Mutual Information And The Correlation Function}

In this section, we prove that the mutual information provides an upper bound to the correlation function. Given two random variables $X$ and $Y$, we start with Pinsker's inequality for mutual information 
\begin{equation}
\begin{split}
    I(X, Y) &= D_{\text{KL}}\left( P_{X Y}(x, y) || P_X(x)P_Y(y) \right)\\
    &\geq \frac{1}{2}\left(\sum_{x\in X, y\in Y}\left\lvert P_{X Y}(x, y) - P_X(x) P_Y(y) \right\rvert\right)^2.
\end{split}
\label{eqn:pinsker}
\end{equation}
Suppose $f(x)$ and $g(y)$ are bounded functions on $X$ and $Y$ respectively, then the following inequality is valid:
\begin{equation}
    0\leq \frac{|f(x)g(y)|}{\|f(x)\|_{\infty}\|g(y)\|_{\infty}} \leq 1\,,
    \label{eqn:bound}
\end{equation}
where $\|\cdot\|_{\infty}$ denotes the supremum norm. We add this fraction as a factor on the right-hand side of Eq.~\eqref{eqn:pinsker} to arrive at
\begin{equation}
\begin{split}
    I(X,Y) &\geq \frac{1}{2}\left(\sum_{x\in X, y\in Y}\biggl\lvert \left[P_{X Y}(x, y)-P_X(x) P_Y(y)\right] \frac{f(x)g(y)}{\|f(x)\|_{\infty}\|g(y)\|_{\infty}}\biggr\rvert\right)^2 \\
    &\geq \frac{1}{2}\left( \frac{\langle f(x)g(y)\rangle - \langle f(x)\rangle \langle g(y)\rangle}{\|f(x)\|_{\infty} \|g(y)\|_{\infty}}\right)^2 .
\end{split}
\label{eqn:mutual_information}
\end{equation}
This is the desired relation between the mutual information and the normalized correlation function of variables $f$ and $g$. 

\section{The Linear RG Loss Function}

For a linear RG transformation $T_\theta$ and the Gaussian actions \eqref{eqn:gaussian_action} and \eqref{eqn:latent_action}, the loss function $L$ in Eq.~\eqref{eqn:7} can be simplified to
\begin{equation}
    \begin{split}
        L &= \displaystyle \mathop{\mathbb{E}}_{z \sim P[z]} \left\Vert \frac{1}{2}\frac{\delta }{\delta z} \left(z^T S_{\text{z}}z - z^T \left(T_\theta^{-1}\right)^T S_{\text{UV}}T_\theta^{-1}z\right)\right\Vert_{2}^{2} \\
        &= \displaystyle \mathop{\mathbb{E}}_{z \sim P[z]} \left\Vert z^T \left(S_{\text{z}} - \left(T_\theta^{-1}\right)^T S_{\text{UV}}T_\theta^{-1}\right)  \right\Vert_{2}^{2} \\
        &= \Tr \left[\left(S_{\text{z}} - \left(T_\theta^{-1}\right)^T S_{\text{UV}}T_\theta^{-1}\right)^2 S_{\text{z}}^{-1}\, \right] .
    \end{split}
    \label{eqn:A1}
\end{equation}
The above loss function reaches its global minimum (equal to zero) if and only if $S_{\text{UV}} = T_\theta^{T} S_{\text{z}}T_\theta$ or equivalently if Eq.~\eqref{eqn:S_z_from_S_UV} holds after a Fourier transformation. 

In the 1D case, our latent configuration $z$ has alternating pattern:
\begin{equation}
    z = (\psi_0, \xi_0, \ldots, \psi_{N - 1}, \xi_{N - 1}).
    \label{eqn:pattern}
\end{equation}
The convolution matrix $T_\theta$ for the RG transformation $z = T_\theta (\phi)$ specified by the kernels \eqref{eqn:C_1_C_2} has the following block circulant form:
\begin{equation}
    T_\theta = \text{bcirc}\left[\begin{pmatrix}
    a & c \\
    0 & f
\end{pmatrix},
\begin{pmatrix}
    d & 0 \\
    g & h
\end{pmatrix},
\ldots
\right]
\end{equation}
(only the nonzero blocks are shown).
Projecting this $T_\theta$ onto the basis of Fourier harmonics,
we obtain Eq.~\eqref{eqn:tilde_T_theta}.
Writing it for each of the four matrix elements, we arrive at four equations specifying the six kernel parameters $a$ through $h$:
\begin{equation}
    \begin{split}
        \left(a+ e^{-i q_n} d\right) \left(a+ e^{i q_n} d\right) (r_{\text{IR}} + 2 - 2 \cos q_n) + g^2 - r_{\text{UV}} - 2 &= 0\,, \\ 
        \left(ac + e^{-i q_n} cd\right) (r_{\text{IR}} + 2 - 2 \cos q_n) + e^{-i q_n} (f g+1) + g h + 1 &= 0\,, \\
        \left(ac + e^{i q_n} cd\right) (r_{\text{IR}} + 2 - 2 \cos q_n) + e^{i q_n} (f g+1) + g h + 1 &= 0\,, \\
        c^2(r_{\text{IR}} + 2 - 2 \cos q_n) + 2 fh \cos q_n + f^2 + h^2 - r_{\text{UV}} - 2 &= 0\,. 
    \end{split}
\end{equation}
Since the above equations hold at all momenta $q_n$, the coefficients in front of each power of $e^{i q_n}$ must sum up to zero. The resulting set of equations has the unique solution reported in the main text. 

\bibliography{RGFlow.bib}

\end{document}